\documentclass[12pt]{article}
\newcommand{\sect}[1]{\setcounter{equation}{0}\section{#1}}

\evensidemargin=\oddsidemargin
\newcommand{\be}{\begin{equation}}
\newcommand{\ee}{\end{equation}}
\newcommand{\bea}{\begin{eqnarray}}
\newcommand{\eea}{\end{eqnarray}}
\newcommand{\p}{\partial}

\newcommand{\nn}{\nonumber}

\begin{document}
\renewcommand{\thefootnote}{\fnsymbol{footnote}}
\begin{titlepage}
\begin{flushright}
GEF-TH-05/2006\\
hep-th/0604172\\
\end{flushright}
\vskip .7in
\begin{center}
{\Large \bf D-Brane solutions in a light-like linear dilaton background}
\vskip .7in
{\large Rashmi R. Nayak$^a$}\footnote{e-mail:{\tt Rashmi.Nayak@ge.infn.it
}}, and {\large Kamal L. Panigrahi$^{a,b}$}
\footnote{e-mail: {\tt Kamal.Panigrahi@ge.infn.it,panigrahi@iitg.ernet.in}}
\vskip .2in
{$^a$ \it Dipartimento di Fisica,
Universita' \& INFN Sezione di Genova, ``Genova'' \\
Via Dodecaneso 33, 16146, Genova, Italy}\\
\vskip .2in
{$^b$ \it Physics Department,
Indian Institute of Technology, Guwahati, 781 039 India}\\
\vspace{.7in}
\begin{abstract}
\vskip .5in \noindent The light-like linear dilaton background
presents a simple time dependent solution of type II supergravity
equations of motion that preserves 1/2 supersymmetry in ten
dimensions. We construct supergravity D-brane solutions in a linear
dilaton background starting from the known intersecting brane
solutions in string theory. By applying a Penrose limit on the
intersecting $(NS1-NS5-NS5')$- brane solution, we find out a D5-brane
in a linear dilaton background. We solve the Killing spinor
equations for the brane solutions explicitly, and show that they
preserve 1/4 supersymmetry. We also find a M5-brane solution in
eleven dimensional supergravity.
\end{abstract}
\end{center}
\vfill

\end{titlepage}
\setcounter{footnote}{0}
\renewcommand{\thefootnote}{\arabic{footnote}}
\tableofcontents \sect{Introduction} Study of time dependent physics
is interesting and challenging. Till now our knowledge of time
dependent phenomena in string theory is quite limited, and hence it
is important to learn more about them. Often they contain space time
singularities, and the challenge is that in order to make definite
statements about the perturbative physics we should learn how to
resolve the singularities. One of the key problems of quantum
gravity is to solve the cosmic singularity, which is widely believed
to be resolved in the framework of string theory. Unlike the
orbifold or the conifold singularity, the big bang singularity is
space-like, and its resolution requires knowledge of string theory
in time dependent or cosmological backgrounds. For example, the time
dependent orbifold models have been constructed to address the
issues of resolution of and physics near the cosmological
singularities \cite{ Liu:2002ft,Liu:2002kb,Cornalba:2002fi}. The
time dependent orbifold models are promising as they are
supersymmetric and there is no null killing vectors. So the problem
of particle production was absent and it was fairly simple to solve
these models. However, it turns out that these models are unstable
due to the large blue shift effect and so on
\cite{Horowitz:2002mw,Lawrence:2002aj,Berkooz:2002je}. There are
several other attempts for studying string theory in time dependent
background with cosmological singularity and some related behavior
in (see for example: \cite{Horowitz:1989bv}-\cite{Hikida:2005ec}).
Another class of time dependent model has been proposed in the form
of the `rolling' of the open string tachyon \cite{Sen:2002nu} on an
unstable brane (brane-anti brane pair). Recently, there are attempts
to replace the cosmological singularity by a closed string tachyon
condensation phase and study the perturbative string
amplitudes \cite{McGreevy:2005ci}. More recently, the ``matrix big
bang'' proposal has been put forth in \cite{Craps:2005wd}. It has
been argued that in a light-like linear dilaton background the
matrix degrees of freedom, rather than the point particles or the
perturbative string states, explain the correct physics near the
big-bang singularity\footnote{see
\cite{Li:2005sz}-\cite{Ishino:2006nx} for related work along this
line.}. In this context, the dual matrix string description has been
given in terms of a 2-dimensional supersymmetric Yang-Mills theory
on a time dependent worldsheet that is the Milne orbifold of a
2-dimensional flat Minkowski space. This background, in particular,
presents a simple time dependent solution of type II supergravity
equations of motion which preserves 1/2 supersymmetry in ten
dimensions. In order to have a complete picture of what is going on,
one should try to learn more about the dual gauge theory of some
class of time dependent solutions with a linear dilaton. It might
help us to understand the physics near the singularity from the dual
gauge theory view point. Supergravity backgrounds often help in
understanding the nature of time dependent sources in gauge theory,
and hence it is important to investigate such solutions in detail.
In view of finding out time dependent solutions in string theory,
and to learn about the physics near the space-like singularity, in
this paper we construct D-brane solutions in a linear dilaton
background. Starting from the intersecting brane solutions of type
$1_{NS}$ + $5_{NS}$ + $5'_{NS}$ in supergravity, we obtain D5-brane
solutions by applying a particular Penrose limit
\cite{Craps:2005wd}. We further analyze the supersymmetric
properties of the solution that we found.

The rest of the paper is organized as follows. In section-2 we make
a quick review of the linear dilaton background that can be obtained
by taking a particular Penrose limit of a stack of NS5-branes in the
near horizon region. Section-3, is devoted to the construction of
D-brane supergravity solutions by the application of a Penrose limit 
on a configuration of intersecting branes in supergravity . 
We also investigate
the geodesic equations, construct a M5-brane solution in eleven
dimensional supergravity. In section-4, we investigate the
space-time supersymmetry of the D5-brane solution. We present our
conclusions in section-5.

\sect{Light-like linear dilaton background as a Penrose limit} It
was pointed out in \cite{Craps:2005wd} that a light-like linear
dilaton can be obtained by taking a particular `Penrose limit' on a
stack of NS5-branes in type II string theory. The supergravity
solution of a stack of NS5-branes can be written in string frame as,
\bea 
ds^2 &=& -dt^2 + dy^2_5 + H(r)\left(dr^2 + r^2 (d\theta^2
+ \cos^2 \theta d\psi^2 + sin^2 \theta d\phi^2)\right)\,, \nn\\
H^{(3)} &=& N \epsilon_3\,,\qquad 
e^{2\Phi} = g^2_s H(r)\,,\>\>\>\>
H(r) = 1 + \frac{N l^2_s}{r^2}\,, 
\eea 
where $H^{(3)}$ is the NS-NS field strength, $\epsilon_3$ is the 
volume form on the transverse $S_3$ and $N$ is the NS5-brane charge.
$H(r)$ is the harmonic function in the transverse space. The
near-horizon geometry corresponds to the limit $r\rightarrow 0$
which removes the $1$ in $H(r)$ and, on rescaling the time $ (t =
\sqrt{N} l_s \tilde{t})$, leads to the following, 
\bea ds^2 &=& N
l^2_s \left( -d\tilde{t}^2 + {{dr^2}\over {r^2}} + \cos^2 \theta
d\psi^2 + d\theta^2 + sin^2 \theta d\phi^2 \right)+
dy^2_5 \ , \nn \\
e^{2\Phi} &=& \frac{Nl^2_s g^2_s}{r^2} \ , 
\eea 
there is also the
three form NS-NS field strength whose explicit form we are not
mentioning here. To take the Penrose limit, one is interested in
boosting along a radial rather than an angular null
geodesic \cite{Hubeny:2002vf}. The starting point is to make the
following replacements 
\bea 
(\tilde t, x)\rightarrow (x^+,
x^-),\>\>\> \tilde t = x^+ - x^-, \>\>\> r = \sqrt{N}l_s e^{x^+}.
\eea 
This gives the metric and the dilaton in the following form
\bea 
ds^2 &=& Nl^2_s\left[2 dx^+ dx^- - (dx^-)^2 + d\theta^2
+ \cos^2 \theta d\psi^2 + \sin^2 \theta d\phi^2 \right] + dy^2_5 \, \nn \\
e^{2\Phi} &=& g^2_s e^{-2 x^+} 
\eea 
Next step is to rescale 
\bea x^-
\rightarrow \frac{x^-}{N}, \>\>\> \theta \rightarrow \frac{\theta}{\sqrt{N}}, 
\>\>\>\> \psi \rightarrow \frac{\psi}{\sqrt{N}} 
\eea 
and take the limit $N\rightarrow \infty$.
This sends the $H^{(3)}\sim N\sin\theta \ d\theta \ d\psi \ d\phi
\rightarrow N^{-1/2} \rightarrow 0$, and finally one is left with a
metric and a dilaton that is linear along $x^+$, 
\bea 
ds^2 = l^2_s\left(2 dx^+
dx^- + \sum^{3}_{i=1} {(dx^i)}^2\right) + \sum^{5}_{a=1} {(dy^a)}^2
, \>\>\>\> \Phi = \Phi_0 - x^+ \ , 
\label{ld-NS5} 
\eea 
which describes the light-like linear dilaton in a flat space-time. It is
fairly straightforward to check the space-time supersymmetry of this
solution, that is given by the condition 
\bea 
\Gamma^{\hat +}\epsilon = 0 
\eea 
where $\epsilon$ represents the space time Killing
spinor in ten dimensions and the hat represents the corresponding 
tangent space index. Hence the background preserves half of the
supersymmetry. In the subsequent analysis, we would like to see how
the supersymmetry changes in the presence of D-branes.

Next, let us review  some of the geometric features of the
background with a linear dilaton. The geometry specified by
(\ref{ld-NS5}), from a string frame view point, contains a
time-dependence in the coupling constant but the metric is flat.
However, to study the geodesic equations for such a background, we
pass on to the Einstein frame. By using the relation 
\bea 
ds^2_{E}= e^{-\Phi /2} ds^2_{string} 
\eea 
we get the metric in the Einstein frame 
\bea 
ds^2_{E} = e^{x^+ /2} \left(l^2_s \left(2 dx^+ dx^- +
\sum^{3}_{i=1} {(dx^i)}^2\right)  + \sum^{5}_{a=1} {(dy^a)}^2\right)
\eea 
Viewed from an Einstein frame, the space time originates at a
big-bang singularity as $x^+ \rightarrow -\infty$, as the scale
factor goes to zero. The geodesic equation for this background in
Einstein frame, along $x^+$ ({\rm for} $(x^-, x^i, y^a) = \rm
{constant})$ is written as 
\bea 
\frac{d^2 x^{\mu}}{d\sigma^2} +
\Gamma^{\mu}_{\alpha\beta} \frac{dx^{\alpha}}{d\sigma}\frac{d
x^{\beta}}{d\sigma} = 0 
\eea 
which for $\Gamma^{+}_{++} = 1/2$, can be given by 
\bea 
\frac{d^2 x^+}{d\sigma^2} +
\frac{1}{2}\left(\frac{dx^+}{d\sigma}\right)^2 = 0. 
\eea 
Hence the
affine parameter is given by 
\bea 
\sigma = e^{x^+ /2}, 
\eea 
upto an affine transformation. So the singularity at $x^+ \rightarrow
-\infty$ correspond to $\sigma \rightarrow 0$. Written in $\sigma$
variable, we get the following metric 
\bea 
ds^2 = l^2_s\left(4
d\sigma dx^- + \sigma \sum^{3}_{i=1} {(dx^i)}^2\right) + \sigma
\sum^{5}_{a=1} {(dy^a)}^2 
\eea 
In this frame, the Riemann tensors
diverge at $\sigma = 0$, thereby showing a curvature singularity. In
other words this gives a divergent tidal force felt by an inertial
observer. In the next section, we would like to see if this behavior
is affected by the introduction of non-perturbative objects like
D-branes.

\sect{Classical D-brane solutions in a light-like linear dilaton
background} To obtain our solution, we now begin by writing the
$1_{NS}$ + $5_{NS}$ + $5'_{NS}$ solution given in
\cite{Papadopoulos:1999tw} with the metric:
\begin{equation}
ds^2 = g_1^{-1}(x,y)(-dt^2 + dz^2) + H_5(x) dx^2_i
+ H'_5(y) dy^2_m, \label{g155}
\end{equation}
supported by the NS-NS field strengths
\begin{equation}
d{\rm B} = dg_1^{-1}\wedge dt \wedge dz + *dH_5 + *dH'_5,
\label{b155}
\end{equation}
and the dilaton
\begin{equation}
e^{2\Phi} = {H_5(x)H'_5(y)\over g_1(x,y)},
\label{d155}
\end{equation}
where
\begin{equation}
[H'_5(y) {\partial_x}^2 + H_5(x) {\partial_y}^2] g_1(x,y)=0.
\end{equation}
A particular solution for $g_1$ is given as:
\begin{equation}
g_1(x,y) = H_1(x)H'_1(y),
\end{equation}
with the harmonic functions given by the following expression
\begin{equation}
H_1 = 1 + {N_1^2l^2_s\over x^2},~~~H_5 = 1 + {N_5^2l^2_s\over x^2},~~~
H'_1 = 1 + {{N'}_1^2l^2_s\over y^2},~~~H'_5 = 1 + {{N'}_5^2l^2_s\over y^2}.
\end{equation}
We would like to note that the D5-branes wrapping AdS$_3\times$ S$^3$
has been obtained in \cite{Papadopoulos:1999tw} by taking a
particular near horizon limit. We also note that by taking a Penrose
limit along a null geodesic, D-brane solutions in the pp-wave background
with R-R three form flux has been obtained in \cite{Kumar:2002ps}. 
However, in the present
paper we are interested in another Penrose limit proposed in
\cite{Craps:2005wd} in order to get D-brane solutions in a
light-like linear dilaton background.

To proceed further, first we would like to set all the string
charges to zero, by putting $H_1=H'_1 = 1$. Then the above solution
becomes a direct product of $NS5 + NS5'$ configuration. Next, we
apply the Penrose limit as \cite{Craps:2005wd,Hubeny:2002vf}. By
rescaling the conventional time $ t = \sqrt{N_5} l_s \tilde{t}$, the
metric above can be written as 
\bea 
ds^2 = N_5 l^2_s \left[-d\tilde
t^2 + \frac{dx^2}{x^2} + d\Omega_3^2\right] + dz^2 + H'_5 dy^2_m
\eea 
As in \cite{Craps:2005wd} we will also be interested in
boosting along a radial null geodesic in this solution. To do this
let's make the following replacement 
\bea 
(\tilde t, x)\rightarrow
(x^+, x^-),\>\>\> \tilde t = x^+ - x^-, \>\>\> x = \sqrt{N_5}l_s
e^{x^+}. 
\eea
and apply the rescaling 
\bea x^- \rightarrow \frac{x^-}{N_5}, \>\>\>
\theta \rightarrow \frac{\theta}{\sqrt{N_5}}, \>\>\>\> \psi
\rightarrow \frac{\psi}{\sqrt{N_5}} 
\eea 
and finally take the limit
$N_5 \rightarrow \infty$. In this limit the three form field
strength correspond to the first NS5-brane $*dH_5 \rightarrow 0$.
Now the metric for the $NS5'$-brane looks like
\begin{eqnarray}
ds^2 &=& \left[2dx^+ dx^- + \sum^{4}_{a=1}dx^2_a\right] +
H'_5\sum^{8}_{m=5}dy^2_m, \>\>\>  H_{mnp} = \epsilon_{mnpq}
\partial_q H'_5
\end{eqnarray}
The dilaton after this rescaling looks like \bea e^{2\Phi} &=&
e^{-2x^+} H'_5. \eea Applying a S-duality\footnote{Under S-duality,
$\Phi\rightarrow -\Phi$, $ds^2_{string} \rightarrow \exp(-\Phi) 
ds^2_{string}$, and the NS-NS fields change to RR fields.} 
we get the following metric, dilaton and the Ramond-Ramond field 
strength $(F_{mnp})$ for the system of D5-branes in a linear 
dilaton background:
\begin{eqnarray}
ds^2 &=& e^{x^+} H'^{-\frac{1}{2}}_5\left[2dx^+ dx^- +
\sum^{4}_{a=1}dx^2_a\right] + e^{x^+}
H'^{\frac{1}{2}}_5\sum^{8}_{m=5}dy^2_m \cr & \cr 
e^{2\Phi} &=& e^{2x^+} H'^{-1}_5, \>\>\>\> 
F_{mnp} = \epsilon_{mnpq}\partial_q H'_5
\label{D5-ld}
\end{eqnarray}
We have further checked that the above solution solve the type IIB
field equations. We interpret the solution as D-brane in a linear
dilaton background. We note that by putting $H'_5 = 1$, we get both
the string frame metric and the coupling constant that are time
dependent.

Other D-brane solutions can be found out by applying T-dualities
along the transverse and worldvolume directions of the above
D5-brane solutions. For example, the D4-brane solution can be
generated in the following way. First one delocalizes the above
D5-brane solution (\ref{D5-ld}) along one of the worldvolume
directions (say $x^4$), and then apply a T-duality along that. The
D4-brane solution can be written by
\begin{eqnarray}
ds^2 &=& e^{x^+} H^{-\frac{1}{2}}_4\left[2dx^+ dx^- + \sum^{3}_{a=1}
dx^2_a\right] + e^{x^+} H^{\frac{1}{2}}_4\sum^{8}_{m=4}dy^2_m \cr &
\cr e^{2\Phi} &=& e^{x^+} H^{-{\frac{1}{2}}}_4, \>\>\>\> F_{mnpq} =
\epsilon_{mnpqr} \partial_r H_4 
\label{D4-ld}
\end{eqnarray}
where $H_4 = 1 + \frac{Q_4}{r^3}$ is the Harmonic function in the
transverse space.

Next we would like to get a M5-brane solution starting with the
above D4-brane solution in the linear dilaton background. Using the
well known relation between the 10d and 11d metric: 
\bea 
ds^2_{11} = e^{-\frac{2\Phi}{3}} ds^2_{10} + e^{\frac{4\Phi}{3}} (dx_{11} +
A_{\mu} dx^{\mu})^2, 
\eea 
where $ds^2_{11}$ and $ds^2_{10}$
represent the metric in eleven and ten dimensions respectively, and
$A_{\mu}$ is the one-form field (which is zero in the present case).
One can easily see that the M5-brane solution is given by 
\bea 
ds^2 &=& e^{\frac{2x^+}{3}} f^{-\frac{1}{3}}\left[2dx^+ dx^- +
\sum^{3}_{a=1}dx^2_a + (dx_{11})^2\right] + e^{\frac{2x^+}{3}}
f^{\frac{2}{3}}\sum^{8}_{m=4}dy^2_m, \cr & \cr 
C^{(4)} &=& F^{(4)}
\eea 
where $F^{(4)}$ is the four form field strength given in eqn.
(\ref{D4-ld}) and $f = 1 + \frac{N l^3_p}{r^3}$, with $l_p$ being the
eleven dimensional Plank length. 

It is easy to see that the solutions presented above contain
singularity at $x^+ \rightarrow - \infty$, as the metric components
goes to zero. It is worthy however to know the geometric properties
of the background.
Let us focus on the geodesic at constant $x^-, x^a$, and at $x^m =0
$, that is the trajectory along $x^+$: 
\bea 
\frac{d^2 x^+}{d\sigma^2} + \left(\frac{d x^+}{d\sigma}\right)^2 = 0
\eea 
which gives 
\bea 
e^{x^+}\frac{d x^+}{d\sigma} = {\rm constant}.
\eea 
Hence the affine parameter is given by 
\bea 
\sigma = e^{x^+}
\eea 
up to an affine transformation. Therefore the singularity at $x^+ \rightarrow
-\infty$ correspond to $\sigma = 0$ and it has finite affine
distance to all points inside.

\sect{Spacetime Supersymmetry}
Next, we would like to check the supersymmetric properties of
the D5-brane presented in (\ref{D5-ld}).
The supersymmetry variation of dilatino and
gravitino fields of type IIB supergravity in ten dimension,
in string frame, is given by \cite{Schwarz:1983qr,Hassan:1999bv}:
\begin{eqnarray}
\delta \lambda_{\pm} &=& {1\over2}(\Gamma^{\lambda}\partial_{\lambda}\Phi \mp
{1\over 12} \Gamma^{\mu \nu \rho}H_{\mu \nu \rho})\epsilon_{\pm} + {1\over
  2}e^{\Phi}(\pm \Gamma^{M}F^{(1)}_{M} + {1\over 12} \Gamma^{\mu \nu
  \rho}F^{(3)}_{\mu \nu \rho})\epsilon_{\mp},\\
\label{dilatino}
\delta {\Psi^{\pm}_{\mu}} &=& \Big[\partial_{\mu} + {1\over 4}(w_{\mu
  \hat a \hat b} \mp {1\over 2} H_{\mu \hat{a}
  \hat{b}})\Gamma^{\hat{a}\hat{b}}\Big]\epsilon_{\pm} \cr
& \cr
&+& {1\over 8}e^{\Phi}\Big[\mp \Gamma^{\lambda}F^{(1)}_{\lambda} - {1\over 3!}
\Gamma^{\mu \nu \rho}F^{(3)}_{\mu \nu \rho} \mp {1\over 2.5!}
\Gamma^{\mu \nu \rho \alpha \beta}F^{(5)}_{\mu \nu \rho \alpha
  \beta}\Big]\Gamma_{\mu}\epsilon_{\mp},
\label{gravitino}
\end{eqnarray}
where we have used $(\mu, \nu ,\rho)$ to describe the ten
dimensional space-time indices, and the hat's to represent the
corresponding tangent space indices. Solving the dilatino variation
for the D5-brane solution presented above, we get the following two
conditions to be obeyed by the Killing spinors 
\bea 
\Gamma^{\hat{+}}
\epsilon_{\pm} = 0, \label{back-D5} \eea and \bea \Gamma^{\hat
m}\epsilon_{\pm} + \frac{1}{3!} \epsilon_{\hat m\hat n\hat p\hat r}
\Gamma^{\hat n\hat p\hat r}\epsilon_{\mp} = 0. 
\label{brane-D5} 
\eea
It is easy to see that the first condition comes purely from the
background, where as the second condition is the usual D5-brane
supersymmetry condition even in the flat space time. We note that
for the dilatino variation to be satisfied we infact need both the
conditions. Now we would like to analyze the gravitino variations
for the solution (\ref{D5-ld}). The gravitino variations give the
following conditions on the spinors 
\bea 
\delta {\Psi^{\pm}_{+}}
&\equiv& \p_{+}\epsilon_{\pm}=0,\>\>\> \delta {\Psi^{\pm}_{-}}
\equiv \p_{-}\epsilon_{\pm}=0,\>\>\> \delta {\Psi^{\pm}_{a}} \equiv
\p_{a}\epsilon_{\pm}=0,\cr & \cr \delta {\Psi^{\pm}_{m}} &\equiv&
\p_{m}\epsilon_{\pm} + \frac{1}{8}\frac{H'_{5, m}}{H'_5}
\epsilon_{\pm}=0.\>\>\> 
\label{gravi-D5} 
\eea In writing down the
above variations, we have made use of the conditions (\ref{back-D5})
and (\ref{brane-D5}). Now the conditions (\ref{back-D5}) and
(\ref{brane-D5}) breaks 1/4 of the spacetime supersymmetry, as we
can integrate out the last equation of (\ref{gravi-D5}) with the
solution given by 
\bea 
\epsilon_{\pm} = \exp\left({-\frac{1}{8} \ln
H'_5}\right)\epsilon^0_{\pm} 
\eea 
with a constant spinor
$\epsilon^0_{\pm}$. So the D5-brane solution presented in
(\ref{D5-ld}) preserves 1/4 supersymmetry.

\sect{Conclusion}

In this paper, we have constructed classical solutions for D-branes
in a light-like linear dilaton background. These brane solutions
have been obtained by taking a particular Penrose limit on an
intersecting brane solution in type II supergravity. We have found
out that starting with the intersecting $NS1-NS5-NS5'$ brane
solution, one can apply the near horizon limit followed by a Penrose
limit along a radial null geodesic to obtain branes in a light-like
linear dilaton background. We further have obtained a M5-brane
solution in eleven dimensional supergravity. The supersymmetry
variations reveal that these branes preserve 1/4 of the full type
IIB space-time supersymmetry. We have also pointed out the geodesic
equations and the nature of singularity at $x^+ \rightarrow
-\infty$. One can possibly try to extend the present analysis to all
the D-branes including the intersecting ones in type II string
theory. In particular, one can try to find out an intersecting
$D1-D5$ brane solution in a light-like or null linear dilaton
background. The near horizon limit of such brane configuration can
be thought of as a deformation of AdS$_3\times$ S$^3$ space time. It
might also help us in understanding the physics near the singularity
from the view point of the underlying gauge theory. The construction
of the corresponding matrix model for these class of branes would be
very interesting. We hope to come back to some of these issues in
near future.

\vskip .2in \noindent {\large\bf{Acknowledgements:}} We would like
to thank A. Kumar for reading the manuscript. We thank S. Siwach for
useful discussions. The work of RRN was supported by INFN. The work
KLP was supported partially by PRIN 2004 - "Studi perturbativi e non
perturbativi in teorie quantistiche dei campi per le interazioni
fondamentali".

\end{document}